\documentclass[conference]{IEEEtran}
\IEEEoverridecommandlockouts
\usepackage{amsmath,amssymb,amsfonts}
\usepackage{algorithmic}
\usepackage{graphicx}
\usepackage{textcomp}
\usepackage{xcolor}
\usepackage{blindtext}
\usepackage[normalem]{ulem}
\def\BibTeX{{\rm B\kern-.05em{\sc i\kern-.025em b}\kern-.08em
    T\kern-.1667em\lower.7ex\hbox{E}\kern-.125emX}}

\usepackage[backend=bibtex,style=ieee]{biblatex}
\usepackage{fancyhdr}
\begin{document}

\title{Federated Learning with Dual Attention for Robust Modulation Classification under Attacks\\
{
}
}
\author{
    \IEEEauthorblockN{Han Zhang\IEEEauthorrefmark{1}, Medhat Elsayed\IEEEauthorrefmark{2},Majid Bavand\IEEEauthorrefmark{2}, Raimundas Gaigalas\IEEEauthorrefmark{2},} \IEEEauthorblockN{Yigit
    Ozcan\IEEEauthorrefmark{2} and Melike Erol-Kantarci\IEEEauthorrefmark{1},\IEEEmembership{ Senior Member, IEEE}}
    
    \IEEEauthorblockA{\IEEEauthorrefmark{1} School of Electrical Engineering and Computer Science, University of Ottawa, Ottawa, Canada}
    
    \IEEEauthorblockA{\IEEEauthorrefmark{2} Ericsson Inc., Ottawa, Canada}
    
    \IEEEauthorblockA{\{hzhan363, melike.erolkantarci\}@uottawa.ca, }\IEEEauthorblockA{\{medhat.elsayed, majid.bavand, raimundas.gaigalas, yigit.ozcan\}@ericsson.com}
}

\maketitle

\thispagestyle{fancy}   
\fancyhead{}                
\lhead{Accepted by 2024 IEEE International Conference on Communications (ICC), \copyright2023 IEEE}
\cfoot{}
\renewcommand{\headrulewidth}{0pt}   
\begin{abstract}
Federated learning (FL) allows distributed participants to train machine learning models in a decentralized manner. It can be used for radio signal classification with multiple receivers due to its benefits in terms of privacy and scalability. However, the existing FL algorithms usually suffer from slow and unstable convergence and are vulnerable to poisoning attacks from malicious participants. In this work, we aim to design a versatile FL framework that simultaneously promotes the performance of the model both in a secure system and under attack. To this end, we leverage attention mechanisms as a defense against attacks in FL and propose a robust FL algorithm by integrating the attention mechanisms into the global model aggregation step. To be more specific, two attention models are combined to calculate the amount of attention cast on each participant. It will then be used to determine the weights of local models during the global aggregation. The proposed algorithm is verified on a real-world dataset and it outperforms existing algorithms, both in secure systems and in systems under data poisoning attacks.
\end{abstract}


\section{Introduction} 
Radio signal classification and modulation recognition is a broadly studied wireless application that aims to classify radio signals and detect the employed modulation methods in an autonomous way. It can be used for radio fault detection, spectrum interference monitoring, and other signal regulatory applications. The main difficulty in radio signal classification problems lies in the high dimensional input data that are usually used. In recent years, the deep neural network (DNN) has been widely applied in this area with promising results, given its ability to capture features and extract manageable reduced-dimension representation \cite{chen2021signet}. However, most of the existing DNN-based signal classification works adopt a centralized training approach, in which all the training samples are collected from local servers and gathered at a global server to train a centralized model \cite{shi2019deep}.

Federated learning (FL) is an emerging machine learning (ML) technique that allows distributed participants to train ML models in a decentralized manner. With the extensive applications of FL, it has become an efficient way to train radio signal classification models with multiple receivers due to its benefits in terms of privacy and scalability \cite{wang2021federated}. In FL-based radio signal classification, training data will be kept locally, and models will be trained at each receiver. Only local model weights will be sent to the global server for global model aggregation. In this way, FL can leverage the computing resources of distributed servers and reduce the risk of data leakage. 

However, there are still some limitations of the existing commonly used FL algorithms while applying them to real applications. On the one hand, FL methods always suffer from slow and unstable convergence, which may lead to poor performance. It shows particularly poor performance under non-independent and identically distributed (non-IID) data distribution and while using realistic, imperfect wireless network data \cite{zhao2018federated}. This can be explained by the fact that heterogeneity of the training data held by the participants and the noise in the dataset can lead to weight divergence, which in turn affects the training process.
On the other hand, FL algorithms are usually vulnerable to poisoning attacks from malicious participants. Due to the inherently distributed structure and privacy protection strategies, participants in FL are more difficult to regulate and easier to attack and manipulate. By poisoning the local dataset of one or more participants, the attackers can send malicious updates to the global model, which may lead to performance degradation in both the global model and all the other local models.

To address the above-mentioned limitations of FL, we propose a more efficient and robust FL algorithm by integrating the attention mechanisms into the global model aggregation. Attention mechanisms are deep learning techniques that can be used to provide an additional focus on a specific component by mimicking cognitive attention\cite{vaswani2017attention}. By applying the attention model to FL, the global model can focus on some particular participants by putting more importance and ignore other participants by putting less importance. In this way, the performance of FL can be improved both in secure environments and under attacks.

Inspired by these thoughts, in this work, we propose a dual-attention model weighted FL-based radio signal classification framework. To be more specific, two attention models are combined to calculate the amount of attention that is cast on each participant, it will then be used to determine the weights of local models during the global aggregation. The effectiveness of the proposed algorithm is verified on a real-world dataset. It shows great versatility and outperforms existing algorithms, both in secure systems and in systems under attack. In the secure system, it reduces the classification loss by 7.4\% compared to the commonly used FL algorithm. In the system under attack, it improves the classification accuracy by 11.5\% compared to the baseline defense method. The proposed FL framework can not only be used for modulation classification but also for other wireless network applications. To the best of our knowledge, this is the first work that exploits the possibility of defending against poisoning attacks in FL by attention-weighted global model aggregation.

The rest of the paper is organized as follows. Section II introduces related works, and Section III explains the system model. Section IV describes the detailed implementation of the proposed dual-attention model-based FL algorithm. Section V gives simulation results, and Section VI concludes the paper.

\section{Related works} 
There have been many existing studies that apply DNN to radio signal classification. In \cite{o2018over}, an in-depth study is conducted to investigate the performance of deep learning-based radio signal classification under different settings. In \cite{huang2019data}, data augmentation is performed to classify the modulation categories of received radio signals with limited training data. Most of these works train DNN models in a centralized manner, which may cause privacy and scalability concerns. In \cite{wang2021federated}, a FL-based modulation type classification is proposed under the condition of imbalanced data and varying noise. However, the majority of such techniques may run the risk of slow convergence and vulnerability to attacks in practical applications. In our work, we consider \cite{wang2021federated} as a baseline and compare it with our proposed algorithm. In addition, other works have studied the attacks and defense methods in the radio signal classification application. \cite{kim2021channel} proposes an adversarial attack against deep learning-based wireless signal classifiers by crafting a common adversarial perturbation. Different from this work, we focus on a more stealthy attack technique that involves data poisoning on FL through malicious participants.

Besides the above, several works have applied attention models to FL. In \cite{wang2020attention}, an attention-weighted federated deep reinforcement learning model is proposed for device-to-device assisted heterogeneous collaborative edge caching. In \cite{zhang2023device}, attention-weighted federated learning is leveraged to train traffic steering models across distributed user equipment. In \cite{zhang2021dual}, a dual attention-based federated learning framework is proposed for traffic prediction. Although these works integrate attention schemes with FL in an effective way, they are more applicable to secure systems and do not apply to systems under attack. Even worse, the introduction of attention schemes may make the system more vulnerable to attacks from malicious participants since they may use tactics to grab more attention during global training. Aside from this, these algorithms may have specific requirements for the application scenario, for example, requiring a large number of participants or only applicable to specific applications. In contrast, our proposed algorithm is versatile in both secure systems and systems under attack and shows good performance in both scenarios.

Other than that, there are also some existing works about potential threats and countermeasures in FL frameworks. In \cite{tolpegin2020data}, a label-flipping data poisoning attack is proposed as a common type of attack caused by malicious participants in FL. In this work, we use a label-flipping attack as our threat model in the system under attack. \cite{blanchard2017machine} proposes multi-Krum as an effective defense mechanism against data poisoning attacks in FL through Byzantine-robust global model aggregation, which we use as a baseline in the attacked system. Despite being able to achieve a defensive effect in some application scenarios, the multi-Krum defense requires prior knowledge about the exact number or upper limit of attackers. In contrast, our proposed algorithm does not require any prior knowledge of attacks, and it performs better than the baseline defense.


\section{System model}
In this paper, we consider a wireless radio signal classification system that consists of one transmitter and $K$ receivers. The transmitter transmits signals with $M$ different modulation types, and the receivers will use DNN models to classify the modulation type of the over-the-air received signals.

\begin{figure}[!t]
\centerline{\includegraphics[width=3.0in]{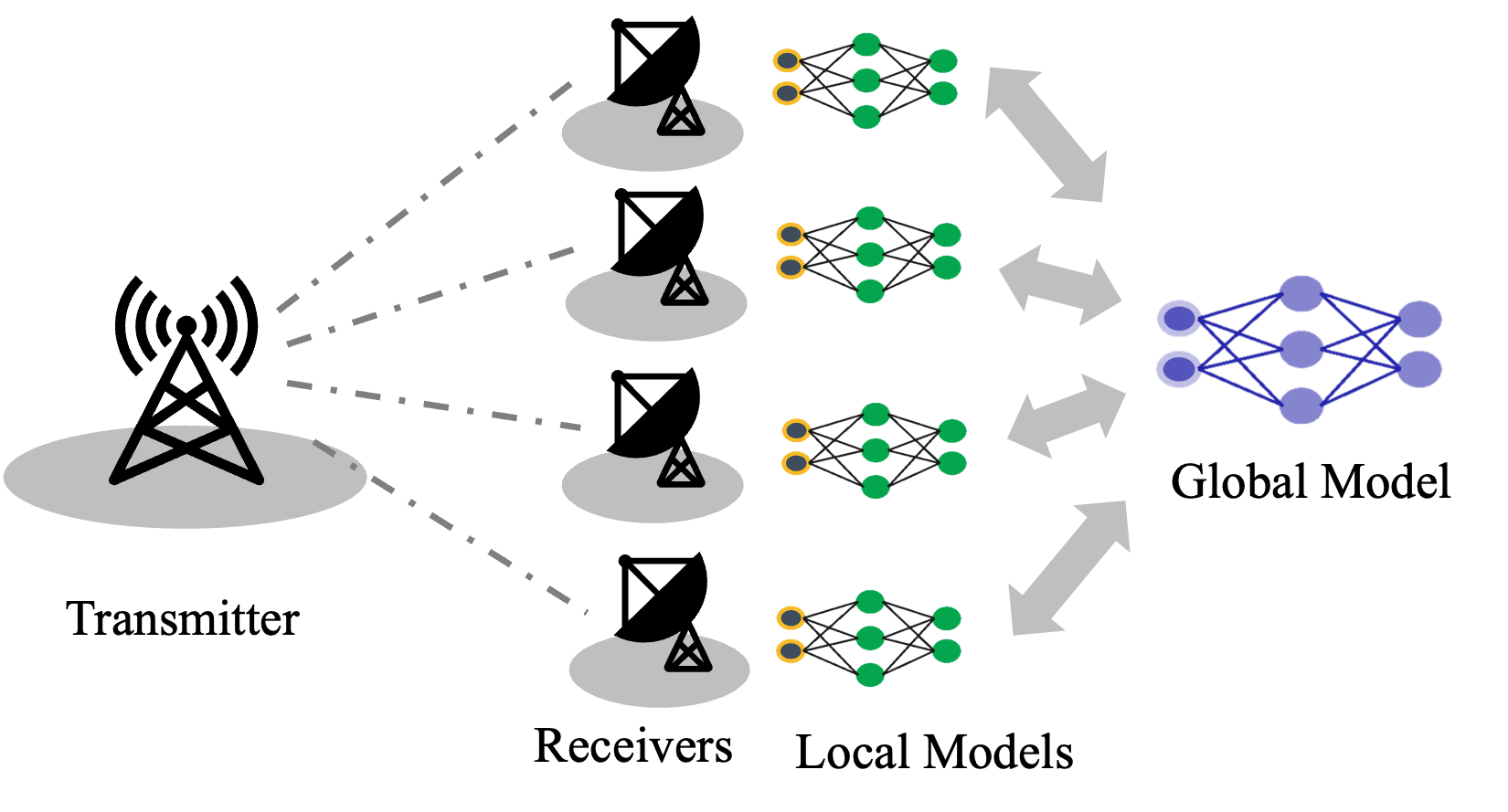}}
\vspace{-10pt}
\caption{System model of FL-enabled radio signal classification.}
\label{fig1}
\vspace{-10pt}
\end{figure}

As shown in Fig. \ref{fig1}, we set up a scenario where multiple receivers work collaboratively to train a model for modulation type classification in a distributed manner. We assume that each receiver collects data on the signal it receives and corresponding modulation types as a local dataset, and uses it to train a local radio signal classification model. After each local training iteration, FL is performed at a centralized server to aggregate the local models into a global model. The objective function of local training at the $k^{th}$ receiver is defined as:
\begin{align}
    L^{Local}_{k}(\omega_k) = -\frac{1}{n_{k}}\sum_{i=1}^{n_{k}}\sum_{m=1}^{M}y_{i,m}log(p_{i,m}),
\end{align}
where $n_{k}$ denotes the number of local training samples at the $k^{th}$ receiver. $\omega_k$ denotes the parameters of the local model at the $k^{th}$ receiver. $p_{i,m}$ denotes the predicted possibility of the DNN model when the modulation type of the $i^{th}$ signal sample is the $m^{th}$ type. $y_{i,m}$ is a binary indicator where $y_{i,m} = 1$ if the modulation type of the $i^{th}$ signal sample is the $m^{th}$ type and 0 otherwise. The local updates can be formulated as follows:
\begin{align}
    \omega_{k,t+1} = \omega_{k,t}-\alpha\frac{\partial L^{Local}_{k}(\omega_k)}{\partial \omega_k},
\end{align}
where $\alpha$ denotes the learning rate of local training.

The optimization goal of FL-enabled radio signal classification can then be defined as:
\begin{align}
    \underset{\omega}{min}\{L^{Global}(\omega) = \sum_{k=1}^{K} L^{Local}_{k}(\omega_k)\},
\end{align}
where $\omega$ denotes the parameters of the local model and $L^{Global}$ denotes the global loss.

The global aggregation can be performed by averaging the model parameters of local models from different receivers, which can be formulated as:
\begin{align}
    \omega_{t+1} = \frac{1}{K}\sum_{k=1}^{K}\omega_{k,t}.
\end{align}

After each iteration of global aggregation, the global model parameters will then be sent back to local models to accelerate the local training. 

In this paper, we consider two scenarios: a secure scenario where all the receivers are benign participants and an attacked scenario where $J$ malicious participants exist. For malicious participants, we perform a label-flipping data poisoning attack and flip the labels of training examples with a source modulation type to a target modulation type while leaving the input signal data unchanged. 

Given the common data distribution pattern in practical scenarios, in this work, we consider a non-IID setting. For each receiver, we assume that there is a dominant modulation type accounting for roughly half of the training samples, while all other modulation types account for the other half of the training data combined. For malicious participants with label-flipping data poisoning attacks, all training samples in the local training dataset that have a dominant class label will change their labels to another different class label. Consequently, during local updates, the malicious participants will learn incorrect updates from poisoned data and spread the incorrect updates to the global model through global aggregation. With these settings, the superiority of the proposed algorithm is demonstrated both in a secure environment and under attack.

\section{Dual attention model-weighted FL for radio signal classification}
To improve the system performance of FL in both secure and attacked scenarios, we propose a dual attention model-weighted FL based on the dot-product attention scheme. In the following subsections, we first explain the core principle of the attention model and then introduce the dual attention model-weighted FL we design for radio signal classification.

\subsection{Attention model}
In this work, scaled dot-product attention is adopted as a fast and space-efficient way to measure the attention weight. The calculation of scaled dot-product attention can be formulated as follows:
\begin{align}
    \eta_k=softmax(\frac{QK^T}{\sqrt{d_k}}),
\end{align}
where $\eta_k$ denotes the attention weight of the $k^{th}$ participant in the attention assessment. $K$ denotes the key, which is the target value that is expected from the participants. $Q$ denotes the queries, which are the actual values that are observed from the participants. $d_k$ denotes the dimension of the key. 

In this way, the closeness between the key and each query is measured by the dot product. If the direction of the query vector is close to the key vector, the dot-product will be large, and the query will gain more attention. On the contrary, if the query vector has an exact opposite direction from the key vector, the dot-product will be small, and the query will gain less attention. $\sqrt{d_k}$ is a scaling vector that can prevent the dot products from growing large due to the high dimension of the key vectors. The final attention weights are given by a SoftMax function to ensure they sum to one.

However, the dot-product attention model is susceptible to the distributions of the queries and the key, thus leading to unstable performance. Therefore, extra steps are designed to make the output of attention values more stable while applying dot-product attention to FL, and the details are described in the next subsection.

\begin{figure}[!t]
\centerline{\includegraphics[width=3.0in]{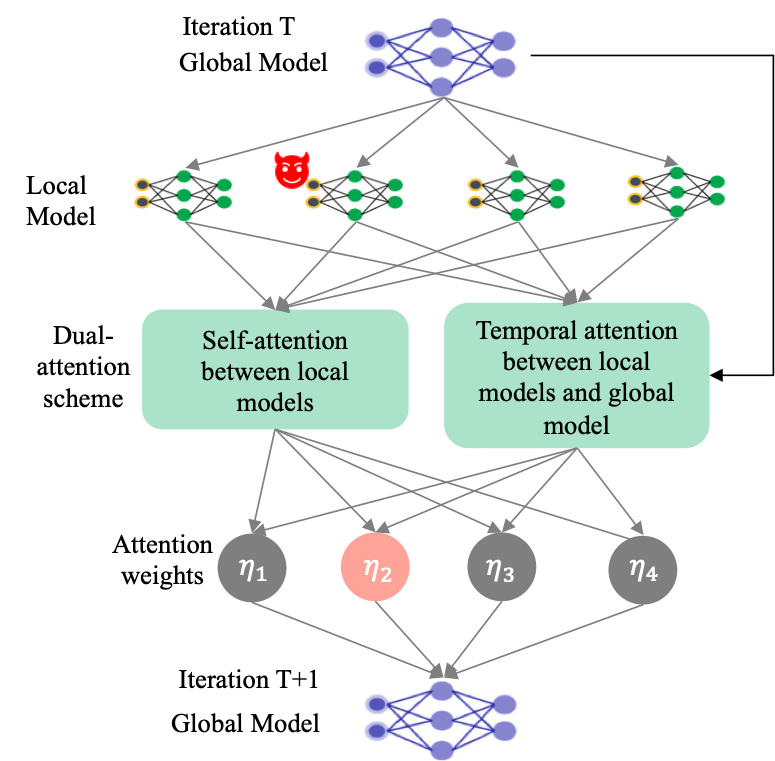}}
\caption{The structure of the proposed dual attention model for FL.}
\label{fig2}
\vspace{-15pt}
\end{figure}

\vspace{-5pt}
\subsection{Dual attention model-weighted FL}

Fig. \ref{fig2} shows the structure of the proposed dual attention model for FL. In this structure, two attention models are set up to evaluate the importance of each local model during the global aggregation. The first attention model is to perform self-attention among all the local models, and the second attention model is to perform temporal attention evaluation between the local models in the current iteration and the global model in the previous iteration. These attention models help to promote FL training by giving different attention weights to different local models. We further illustrate the scheme with a scenario of a system under attack. As it can be observed from Fig. \ref{fig2}, the second participant is assumed to be attacked, and the attack will be reflected in the local model parameters. As a result, the dual-attention scheme will cast less attention on the second local model and give it lower weight. In this way, the incorrect updates from the malicious participants are prevented from affecting the global model. The details of the two attention models are explained as follows.

In the first attention model, self-attention calculation among all the local models is performed to decide how much attention each participant will cast on other participants. This helps to evaluate the connections between local models and to cast more attention to local models that have stronger connections with other local models. For each participant, the local model parameters will be used as the key, and the local model parameters of other local models will be used as the queries. The attention value between the $k^{th}$ participant and the $l^{th}$ participant can be first formulated as the dot product of local parameters of the $k^{th}$ local model and the $l^{th}$ local model:
\begin{align}
    \epsilon_{k,l} = 
    \begin{cases}
        &\frac{\omega_k \omega_k^T}{|\omega_k|.|\omega_k^T|},\ \ \ \ if\ k \neq l \\
        & -inf,\ \ \ \ \ \ \ if\ k = l
    \end{cases}
\end{align}
where $|\omega_k|$ and $|\omega_k^T|$ denote the modulus of $\omega_k$ and $\omega_k^T$.
$inf$ denotes the infinity.

Next, the self-attention value matrix is normalized with mean and variance to ensure the attention model has roughly the same degree of influence on participants in the global aggregation of each iteration. The process can be given as:
\begin{align}
    \hat{\epsilon_{k,l}}= \frac{\epsilon_{k,l}-\mu_{\epsilon}}{\sigma_{\epsilon}},
\end{align}
where $\mu_{\epsilon}$ denotes the mean value of the self-attention value matrix and $\sigma_{\epsilon}$ denotes the variance. 

Then, the attention weight each participant casts on other participants can be calculated through a SoftMax function, which can be formulated as:
\begin{align}
    \eta_{k,l} = \frac{e^{\hat{\epsilon_{k,l}}}}{\sum_{l=1}^{K} e^{\hat{\epsilon_{k,l}}}}.
\end{align}

Finally, the attention weights each participant gets from other clients are summed as the attention weights from the first attention model, which can be formulated as follows:
\begin{align}
    \eta_{k1} = \frac{\sum_{K}^{k=1}\eta_{k,l}}{\sum_{K}^{l=1}\sum_{K}^{k=1}\eta_{k,l}}. 
\end{align}

In the second attention model, similar steps are performed to evaluate how much attention the global model in the previous iteration will cast on the local models from all the participants in the current iteration. In this model, the global model parameters are used as the key, and the local model parameters are used as queries. The attention value the $k^{th}$ participant gets from the second attention model can be calculated as:
\begin{align}
    \epsilon_{k2} = \frac{w_t w_{k,t+1}^T}{|w_t|.|w_{k,t+1}^T|},
\end{align}
where $w_t$ denotes the global model parameters at the $t^{th}$ iteration. Next, the attention values are normalized as:
\begin{align}
    \hat{\epsilon_{k2}}= \frac{\epsilon_{k1}-\mu_{\epsilon}}{\sigma_{\epsilon}}.
\end{align}
The attention weight of the second attention model can then be calculated as follows:
\begin{align}
    \eta_{k2} = \frac{\eta_{k2}}{\sum_{K}^{k2=1}\eta_{k2}}.
\end{align}

After obtaining attention weights from two attention models, the final attention weights for each client can be calculated by adding these two attention weights, which can be formulated as:
\begin{align}
    \eta_k = \beta \eta_{k1} + (1-\beta) \eta_{k2},
\end{align}
where $\beta$ is a coefficient used to balance the ratio of two forms of attention weights. The global aggregation can be then re-formulated as the attention-weighted sum of the local model parameters, which can be given as:
\begin{align}
    \omega_{t+1} = \sum_{k=1}^{K}\eta_{k}\omega_{k,t}.
\end{align}

\section{Numeric Results}
\subsection{Simulation settings}
In this work, we use a real radio signal classification data set generated in \cite{o2018over}. There are $M=11$ modulation types, and we assume there are $K=11$ receivers. The local data set of each receiver is dominated by one receiver. The DNN model used to train the dataset is composed of two convolutional layers, with the number of convolutional kernels being 256 and 80 and the dimensions of the convolutional kernels being $(1,3)$ and $(2,3)$, respectively. The convolutional layers are connected with two dense layers, with the number of neurons being 256 and 11, and ended by a SoftMax layer. The batch size during training is set as 32 and the learning rate is set as 0.0005. We selected data with a signal-to-noise ratio (SNR) between 2 dB and 18 dB for model training and prediction, which contains a total of 99,000 samples. In this dataset, 59,400 samples are used as the training and validation set, and 39,600 samples are used as the test set to evaluate the performance of algorithms. We set $\beta = 0.75$ during the simulation.

In the secure system, we compare the proposed dual-attention model-weighted FL algorithm with the classic FedAvg algorithm proposed in \cite{wang2021federated}\cite{mcmahan2017communication}. In the system under attack, we use multi-Krum proposed in \cite{blanchard2017machine} as the baseline to evaluate the defective effect of proposed algorithm. We assume the best situation for the multi-Krum baseline, which is that the defender knows the exact number of attackers, which is hard to achieve in real situations. Even under such settings, we still demonstrate that our algorithm is superior to the baseline, while it does not require prior knowledge of the attack.
\vspace{-5pt}
\subsection{Simulation results}
In this subsection, we first evaluate the performance of the proposed dual attention model-weighted FL algorithm in a secure system environment and then evaluate it in the presence of attacks on the system.

\begin{figure}[h]
\centerline{\includegraphics[width=2.9in]{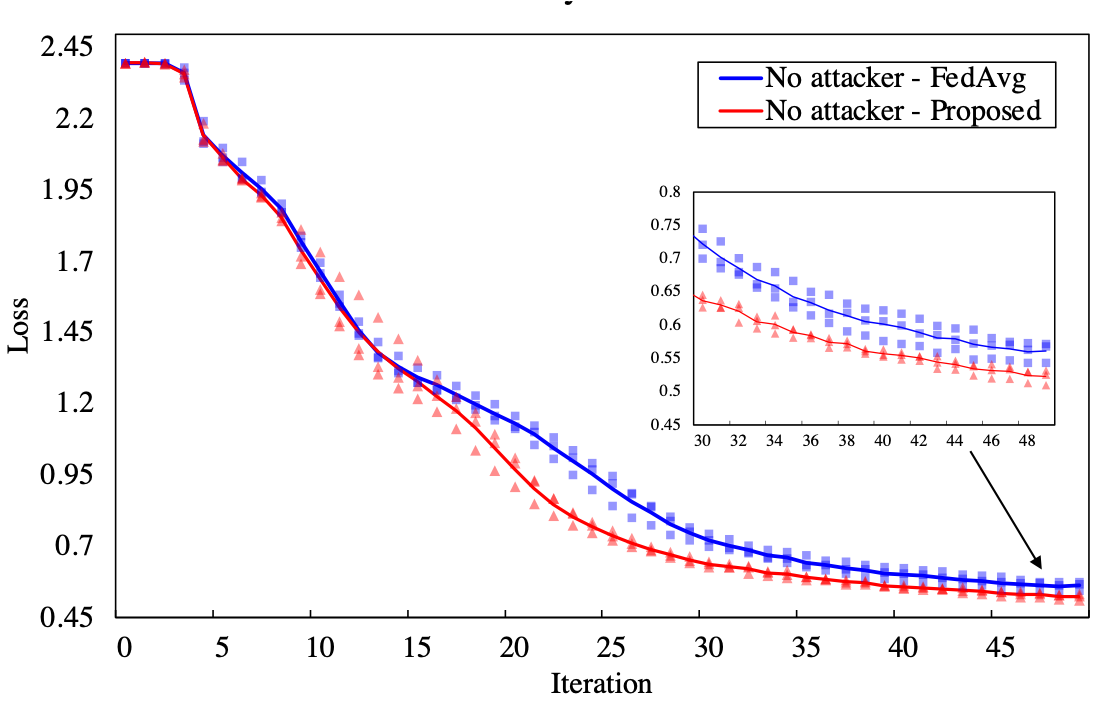}}
\vspace{-10pt}
\caption{The convergence curves of FedAvg and proposed dual attention model-weighted FL in the secure system.}
\label{fig3}
\vspace{-10pt}
\end{figure}
 
Fig. \ref{fig3} shows the convergence curves of FedAvg and the proposed algorithm in the secure system with no attacker. It can be observed that during the training process, the loss of both algorithms gradually decreases and eventually reaches a state of convergence. The loss of our proposed algorithm is always lower than the FedAvg algorithm during the training process, which demonstrates our proposed algorithm converges in a faster and more stable way. This is because the integration of the attention mechanism in the FL algorithm can help to identify the connections between participants and enhance cooperation. After the training, the classification loss of our proposed algorithm is 7.4\% lower than the FedAvg algorithm.

\begin{figure}[h]
\centerline{\includegraphics[width=2.9in]{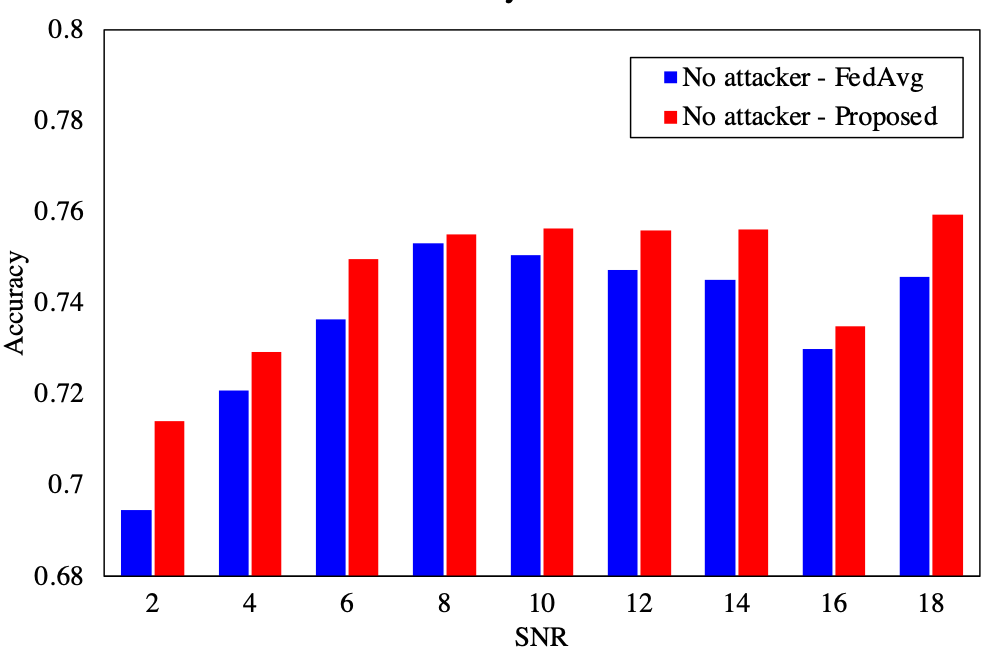}}
\vspace{-10pt}
\caption{The accuracy of modulation type classification under different SNRs in the secure system.}
\label{fig4}
\vspace{-5pt}
\end{figure}

Fig. \ref{fig4} shows the accuracy of modulation type classification under different SNRs in the secure system. It can be observed that the proposed algorithm can have higher classification accuracy than FedAvg under all the SNR settings. This demonstrates that the integration of the attention mechanism in the FL algorithm can help to improve the performance of FL. It is worth noting that when $SNR=16$, an unusual drop in accuracy occurs. The trend is observed in both the baseline algorithm and our proposed algorithm, which suggests that the trend can be attributed to the original dataset as similar trends are also seen in \cite{blanchard2017machine}.

\begin{figure}[h]
\centerline{\includegraphics[width=2.9in]{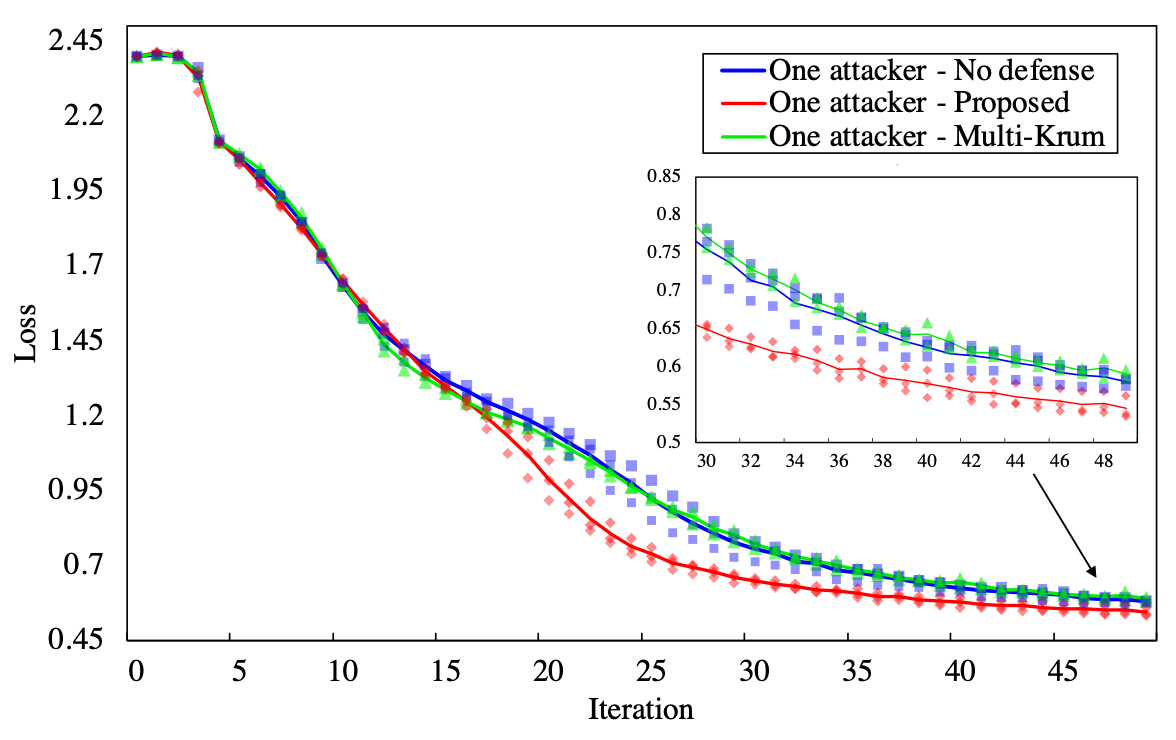}}
\caption{The convergence curves of Multi-Krum and the proposed dual attention model-weighted FL in the presence of one attacker.}
\label{fig5}
\vspace{-10pt}
\end{figure}

\begin{figure}[h]
\centerline{\includegraphics[width=2.9in]{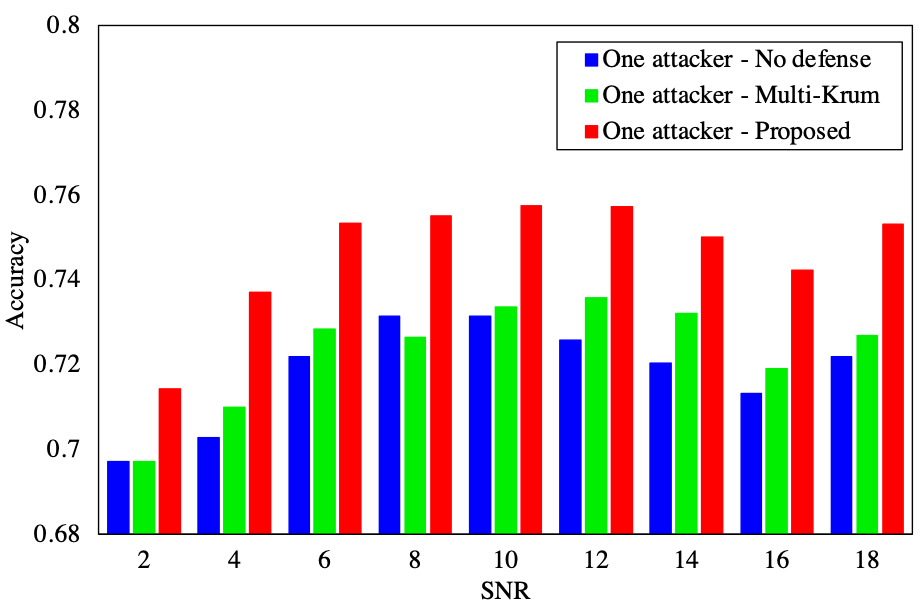}}
\vspace{-10pt}
\caption{The accuracy of modulation type classification under different SNRs in the presence of one attacker.}
\label{fig6}
\vspace{-10pt}
\end{figure}

Then, we perform label-flipping data poisoning attacks on the system and evaluate the robustness of the proposed dual attention model-weighted FL algorithm under attacks. We first set one malicious participant out of 11 participants and flip the labels of a receiver from the 'GFSK' type into the 'CPFSK' type. Fig. \ref{fig5} shows the convergence curves of Multi-Krum and proposed dual attention model-weighted FL, and Fig. \ref{fig6} shows the accuracy of modulation type classification under different SNRs in the presence of one attacker. We also add the convergence curve and the accuracy when there is no defense mechanism as a comparison. It can be observed that in the case of few participants and non-IID training data, multi-Krum has a limited defensive effect. In contrast, our proposed algorithm can still demonstrate a fast and stable convergence in the presence of one attacker. After the training, the classification loss of our proposed algorithm is 10.2\% lower than the multi-Krum algorithm. It also achieves much higher accuracy under different SNRs compared with the multi-Krum defense. With the attention schemes, there is no decrease in the accuracy of classification compared to the secure system.

\begin{figure}[h]
\centerline{\includegraphics[width=2.9in]{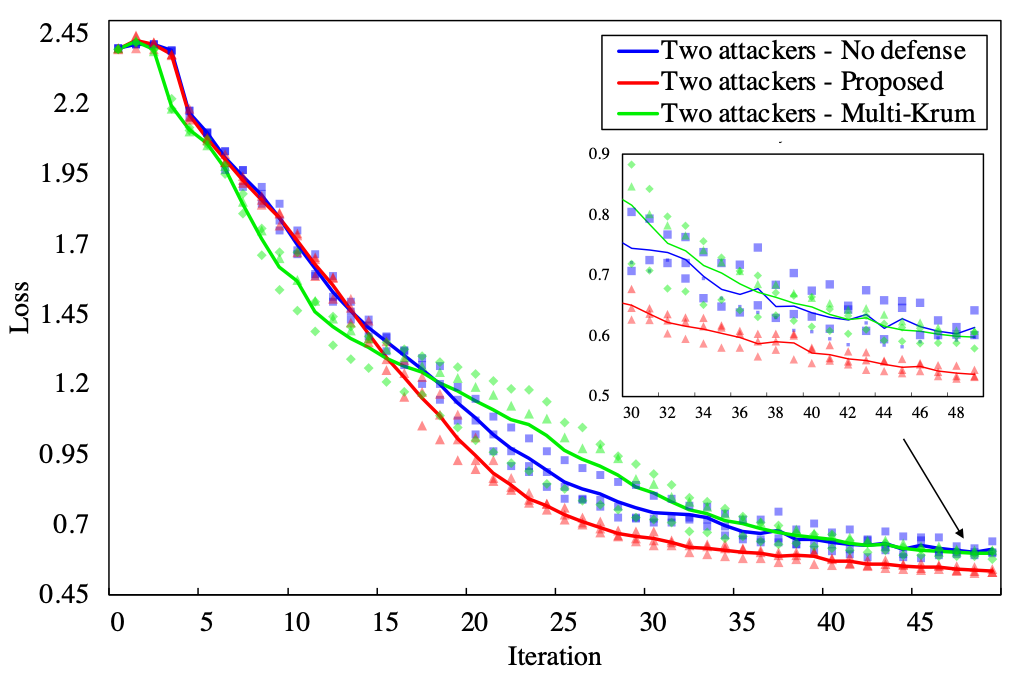}}
\caption{The convergence curves of Multi-Krum and the proposed dual attention model-weighted FL in the presence of two attackers.}
\label{fig7}
\vspace{-10pt}
\end{figure}

\begin{figure}[h]
\centerline{\includegraphics[width=2.9in]{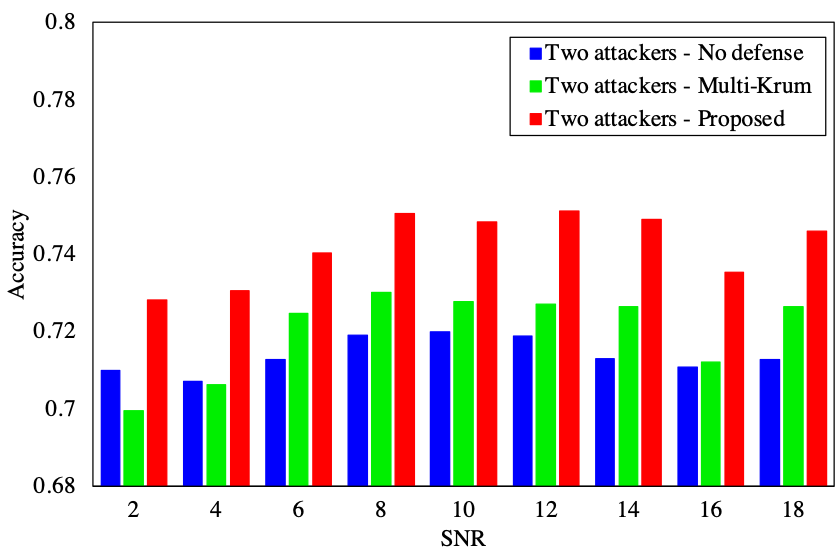}}
\caption{The accuracy of modulation type classification under different SNRs in the presence of two attackers.}
\label{fig8}
\vspace{-15pt}
\end{figure}

 Next, we increase the number of malicious participants to 2 and flip the labels of another receiver from the 'QPSK' type into the 'CPFSK' type. It is worth noting that our proposed algorithm is still applicable in the case of more attackers. However, considering the limitations of attackers' capabilities in real application scenarios, we only consider the case of up to two attackers in our experiments. Fig. \ref{fig7} shows the convergence curves of Multi-Krum and proposed dual attention model-weighted FL, and Fig. \ref{fig8} shows the accuracy of modulation type classification under different SNRs in the presence of two attackers. It can be observed that the Multi-Krum defense has an even worse convergence compared with the system with no defense. This is because the Multi-Krum defense cannot accurately identify malicious participants in every iteration, and the misjudgment in each iteration will cause the model to lose part of the correct training data, which in turn will result in worse performance. In contrast, our proposed algorithm is still able to converge stably and fast and demonstrates almost the same level of accuracy as under a secure system. After the training, the classification loss of our proposed algorithm is 11.5\% lower than the multi-Krum algorithm. These results further validate the defensive effect of the proposed algorithm.

\section{Conclusion}

In this work, we proposed an efficient and robust attention-weighted FL algorithm and applied it to a radio signal classification scenario. We decide on the attention weight for each participant in FL through two attention models. The first attention model performs a self-attention evaluation between all the participants to find the connections between local models. The second attention model decides the temporal attention between the global model in the last iteration and the local models in the current iteration, and it helps to align the training level of participants. The effectiveness of the proposed algorithm is verified on a real-world dataset. It shows great versatility and outperforms existing algorithms, both in secure systems and in systems under attack.  It reduces the classification loss by 7.4\% in the secure system and improves the classification accuracy by 11.5\% under attacks. 

\section*{Acknowledgement}
 This work has been supported by MITACS and Ericsson Canada.

\normalem
\begin{refcontext}[sorting = none]
\small
\printbibliography
\end{refcontext}

\end{document}